\documentclass[preprint,aps,prd,showpacs,nofootinbib,superscriptaddress,tightenlines]{revtex4}

\usepackage{amsfonts}
\usepackage{mathrsfs}
\usepackage{graphicx}
\usepackage{amsmath}
\usepackage{amssymb}
\usepackage{bm}
\usepackage{bbm}
\usepackage{multirow}
\usepackage{alltt}
\usepackage{fancyvrb}
\usepackage[usenames]{color}
\usepackage[T1]{fontenc}

\def\PA#1{\left( #1 \right)}
\def\PB#1{\left[ #1 \right]}
\def\PC#1{\left\{ #1 \right\}}

\def\nn{\nonumber}
\def\Apart{\texttt{Apart} }
\def\APart{\texttt{\$Apart} }

\begin{document}

\title{\texttt{\$Apart}: A Generalized \textsc{Mathematica} \texttt{Apart} Function}

\author{Feng Feng\footnote{E-mail: fengf@ihep.ac.cn} }
\affiliation{Center for High Energy Physics, Peking University, Beijing 100871, China\vspace{0.2cm}}

\date{\today}

\begin{abstract}
We have generalized the \textsc{Mathematica} function \Apart from 1 to $N$ dimension, the generalized function \APart can decompose any linear dependent elements in $\mathcal{V}_{x}^*$ to irreducible ones. The elements in $\mathcal{V}_{x}^*$ can be viewed as the corresponding propagators which involve loop momenta, and the decomposition will be useful when one tries to perform the loop calculations using the packages such as \textsc{Fire} and \textsc{Reduze}, which have implemented the integration by parts (IBP) identities and Lorentz invariance (LI) identities. A description on how to use this package, combined with \textsc{Fire}, \textsc{FeynArts} and \textsc{FeynCalc} packages, to do the one-loop calculations in double quarkonium production in $e^+e^-$ colliders is given, and the full source code for a specific process $(e^+e^-\to J/\psi + \eta_c)$ is also available.
\end{abstract}


\maketitle

{\bf PROGRAM SUMMARY}
\begin{itemize}

\item[]{\em Title of program:} {\tt \$Apart}
\item[]{\em Programming language:} {\tt \textsc{Mathematica}}
\item[]{\em Available from:}  {\tt \verb=http://power.itp.ac.cn/~fengfeng/apart/=}
\item[]{\em Computer:}  Any computer where the \textsc{Mathematica} is running.
\item[]{\em Operating system:} Any capable of running Mathematica.
\item[]{\em External routines:} \textsc{FeynCalc}, \textsc{FeynArts}, \textsc{Fire}
\item[]{\em Keywords:} Next-to-Leading Order (NLO), Integrate By Parts (IBP), Apart
\item[]{\em Classification:} {\tt 11.1}
\item[]{\em Nature of physical problem:}
The traditional method to compute cross sections for a physical process in perturbative quantum field theory involves generating
the amplitudes via Feynman diagrams and performing the dimensionally regularized loop integrals~\cite{'tHooft:1972fi}. Simplifications of the expressions are performed at the analytical level; there, an essential part is the reduction of these loop integrals to a small number of standard integrals. This step can be performed at the amplitude level for tensor integrals or, after contraction of Lorentz indices, at the level of interferences for scalar integrals. Considering the case of scalar integrals, integration by parts (IBP) identities~\cite{Tkachov:1981wb,Chetyrkin:1981qh} and Lorentz invariance (LI) identities~\cite{Gehrmann:1999as} may be used for a systematic reduction to a set of independent integrals, called master integrals (MI). The standard reduction algorithm by Laporta~\cite{laporta} defines an ordering for Feynman integrals, generates identities and solves the resulting system of linear equations. Alternative methods to exploit IBP and LI identities for reductions have been proposed~\cite{Smirnov:2005ky,Smirnov:2006tz,Gluza:2010ws,Schabinger:2011dz}; see also \cite{Lee:2008tj,Grozin:2011mt} and references therein.
Public implementations of different reduction algorithms are available with the computer programs AIR~\cite{Anastasiou:2004vj}, FIRE~\cite{Smirnov:2008iw} and \textsc{Reduze}~\cite{Studerus:2009ye}. The usage of \textsc{Fire}\cite{Smirnov:2008iw}, \textsc{Reduze}\cite{Studerus:2009ye}, etc.~requires that the propagators must be decomposed to independent ones, for 1 dimension, there is a \textsc{Mathematica} function \Apart to do this, while for $N$ dimension there is no such package yet, so we want to generalize the \textsc{Mathematica} function \Apart to \APart in $N$ dimension.
\item[]{\em Method of solution:} We first prove that any linear dependent elements in $\mathcal{V}_{x}^*$ can be decomposed into the summation of linear independent ones, the procedure of the proof gives us a method to perform the decomposition, {\tt \$Apart} is such an \textsc{Mathematica} package that implements this method and generalizes the \textsc{Mathematica} \texttt{Apart} function from 1 to $N$ dimension.
\item[]{\em Running time:} Depends on the complexity of the system.

\end{itemize}

\newpage

\hspace{1pc}
{\bf LONG WRITE-UP}

\section{Introductions and Notations}
A polynomial is a mathematical expression involving a sum of powers in one or more variables multiplied by coefficients. A polynomial in one variable (i.e., a univariate polynomial) with constant coefficients is given by
\begin{equation}
a_n x^n + \cdots + a_2 x^2 + a_1 x + a_0 \;.
\end{equation}
The highest power in a univariate polynomial is called its order, or sometimes its degree, and for a polynomial with more than one variable, i.e. a multivariate polynomial, one needs to find the degree of each term by adding the exponents of each variable in the term, and the largest such degree is the degree of the multivariate polynomial. For a rational expression, i.e. an expression that is the ratio of two polynomials, we can work out its degree by taking the degree of the top (numerator) and subtracting the degree of the bottom (denominator), for example,
\begin{equation}
{\rm Deg}\!\!\PB{\frac{x^3+4x+9}{x^5+2x+1}} \equiv {\rm Deg}\!\!\PB{x^3+4x+9}- {\rm Deg}\!\!\PB{x^5+2x+1} = -2 \;.
\end{equation}

It is well known that there is a function in \textsc{Mathematica} named \Apart which rewrites a rational expression of a univariate polynomial as a sum of terms with minimal denominators and gives the partial fraction decomposition of the rational expression, for example,
\begin{subequations}

\begin{equation}
\texttt{Apart}\!\!\PB{\frac{1}{(x-a) (x-b)}} = \frac{1}{(b-a){x-b}}+\frac{1}{(a-b)(x-a)} \;,
\end{equation}

\begin{eqnarray}
\texttt{Apart}\!\!\PB{\frac{1}{(x-a) (x-b) (x-c)}} &=& -\frac {1} {(a - b) (b - c) (x - b)}+ \frac {1} {(a - c) (b - c) (x - c)} \nn\\
 &&+ \frac {1} {(a - b) (a - c) (x - a)} \;.
\end{eqnarray}

\end{subequations}
There is no such similar function for bivariate polynomials yet; for example, \Apart does not change the following form at all:
\begin{equation}
\Apart\!\!\PB{\frac{1}{x (x + a) (x + y + b)}} = \frac{1}{x (x + a) (x + y + b)}.
\end{equation}
We want to generalize this function to some specific rational expressions of n-variate polynomials; for example, we expect
\begin{equation}\label{expect}
\Apart\!\!\PB{\frac{1}{x (x + a) (x + y + b)},\PC{x,y}} = \frac{1}{a x (x + y + b)} - \frac{1}{a (x + a) (x + y + b)} \;.
\end{equation}

First we introduce the notation. We consider the n-variate polynomial $\mathcal{V}_{x}^*$ with degree less than or equal to 1. The n variables are denoted as $\{ x_i \}_{1\le i \le n}$, and the linear space which is spanned by $n$ independent vectors $\{ x_i \}$ over the coefficient field $\mathcal{F}$ is denoted as $\mathcal{V}_{x}$. It is clear that
\begin{equation}
\mathcal{V}_{x}^* = \mathcal{V}_{x} \oplus \mathcal{F} \;.
\end{equation}

We call the $k$ elements $\PC{e_i = v_i + f_i}_{(1\le i\le k)} \subset \mathcal{V}_{x}^*$ are linear independent if and only if their projective parts in $\mathcal{V}_{x}$, i.e. $\PC{v_i}_{(1\le i\le k)} \subset \mathcal{V}_{x}$, are linear independent.

Now we consider the following special terms generated from the rational operations on the polynomial $\mathcal{V}_{x}^*$:
\begin{equation}\label{General}
\prod_{i=1}^{N} e_i^{n_i}, \qquad  e_i \!\in\! \mathcal{V}_{x}^* \land n_i \!\in\! \mathbb{Z} \land N \!\ge\!1 \;,
\end{equation}
where $\mathbb{Z}$ is the integer set, and $\PC{e_i}_{1\le i\le N}$ are generally not linear independent, i.e. reducible. We want to decompose them into a summation of linear independent, i.e. irreducible, ones:
\begin{equation}\label{Decompostion}
\prod_{i=1}^{N} e_i^{n_i} = \sum_{j} f_j \prod_{i=1}^{N_j} e_{k_{ji}}^{n_{ji}} \;,
\end{equation}
where $1\le k_{ji} \le N$, $1\le N_{j} \le N$ and for any fixed $j$, the $N_j$ elements $\PC{e_{k_{ji}}}_{1\le i\le N_j}$ are linear independent, i.e.~some elements in $\PC{e_i}_{1\le i\le N}$ have been eliminated such that the remaining elements become linear independent.

To give a proof of the decomposition in Eq.~(\ref{Decompostion}), let us consider a special case:
\begin{equation}\label{Special}
F\PA{n_1,\cdots,n_N} \equiv \prod_{i=1}^{N} e_i^{n_i} \;,
\end{equation}
where any $\PA{N-1}$ elements from $\PC{e_i}_{1\le i \le N}$ are linear independent, but the $N$ elements $\PC{e_i}_{1\le i \le N}$ are not, so there exists $\PC{f_i}_{1\le i \le N}$ with all $f_i \ne 0$, such that
\begin{equation}
\sum_{i=1}^N f_i e_i = f \;.
\end{equation}
Note that the linear independency in $\mathcal{V}_{x}^*$ is up to some constant $f$ in the field $\mathcal{F}$.

We first look at Eq.~(\ref{Special}) with all $n_i = -1$, i.e.,
\begin{equation}\label{Simple}
F(-1,\cdots,-1) = \frac{1}{e_1} \frac{1}{e_2} \cdots \frac{1}{e_{N-1}} \frac{1}{e_N}
\end{equation}

\begin{itemize}

\item If $f \ne 0$, we can write Eq.~(\ref{Simple}) as
\begin{eqnarray}
\prod_{i=1}^{N} \frac{1}{e_i} = \frac{f}{f} \prod_{i=1}^{N} \frac{1}{e_i} = \frac{1}{f} \PA{\sum_{j=1}^N f_i e_i} \prod_{i=1}^{N} \frac{1}{e_i}
= \sum_{j=1}^N \frac{f_j}{f} \prod_{i=1,i\ne j}^{N} \frac{1}{e_i} \;,
\end{eqnarray}
since any $\PA{N-1}$ elements from $\PC{e_i}_{1\le i \le N}$ are linear independent, the final expression is irreducible, and the desired decomposition.

\item If $f = 0$, since all $f_i \ne 0$, without loss of generality, we take $f_1$ as an example:
\begin{eqnarray}
\prod_{i=1}^{N} \frac{1}{e_i} = \frac{e_1}{e_1} \prod_{i=1}^{N} \frac{1}{e_i} = \frac{1}{e_1} \PA{-\frac{1}{f_1} \sum_{j=2}^N f_i e_i} \prod_{i=1}^{N} \frac{1}{e_i}
= -\sum_{j=2}^N \frac{f_j}{f_1} \frac{1}{e_1^2} \prod_{i=2,i\ne j}^{N} \frac{1}{e_i} \;.
\end{eqnarray}
We know any $\PA{N-2}$ elements from $\PC{e_i}_{2\le i \le N}$ combined with $e_1$ are linear independent, so the final expression is also irreducible.

\end{itemize}
So we get the decomposition for Eq.~(\ref{Simple}). Now considering the expression of Eq.~(\ref{Special}) with all exponents of $\PC{e_i}$ negative,
\begin{equation}\label{SimpleN}
G\PA{n_1,\cdots,n_N} \equiv F\PA{-n_1,\cdots,-n_N} = \prod_{i=1}^{N} e_i^{-n_i} = \prod_{i=1}^{N} \frac{1}{e_i^{n_i}} \,, \quad n_i > 0 \;,
\end{equation}
we can factorize out a term $\displaystyle\prod_{i=0}^{N} \frac{1}{e_i}$ and perform the decomposition on it as follows:
\begin{itemize}

\item For the case when $f \ne 0$, we have
\begin{eqnarray}
\prod_{i=1}^{N} \frac{1}{e_i^{n_i}} &=& \prod_{k=1}^{N} \frac{1}{e_k^{n_k-1}} \; \prod_{i=1}^{N} \frac{1}{e_i}
= \prod_{k=1}^{N} \frac{1}{e_k^{n_k-1}} \; \PA{ \sum_{j=1}^N \frac{f_j}{f} \prod_{i=1,i\ne j}^{N} \frac{1}{e_i} } \nn\\
&=& \sum_{j=1}^N \frac{f_j}{f} \frac{1}{e_j^{n_j-1}} \prod_{k=1, k\ne j}^{N} \frac{1}{e_k^{n_k}} \;,
\end{eqnarray}
where it is clear that we have decomposed the original term into $N$ terms, and furthermore that these terms have the same form as the original one except that one of the exponents $n_i$ decreases by $1$ in each term, i.e., we get the following recursive relation:
\begin{equation}
G\PA{n_1,\cdots,n_N} = \sum_{j=1}^N \frac{f_j}{f} G\PA{n_1,\cdots, n_j-1,\cdots,n_N}
\end{equation}
and we can repeat the decomposition until one of $n_i$ decreases to $0$.
%

\item When $f = 0$,
\begin{eqnarray}
\prod_{i=1}^{N} \frac{1}{e_i^{n_i}} &=& \prod_{k=1}^{N} \frac{1}{e_k^{n_k-1}} \; \prod_{i=1}^{N} \frac{1}{e_i}
= \prod_{k=1}^{N} \frac{1}{e_k^{n_k-1}} \; \PA{ -\sum_{j=2}^N \frac{f_j}{f_1} \frac{1}{e_1^2} \prod_{i=2,i\ne j}^{N} \frac{1}{e_i} } \nn\\
&=& - \sum_{j=2}^N \frac{f_j}{f_1 e_1} \frac{1}{e_j^{n_j-1}} \prod_{k=1,i\ne j}^{N} \frac{1}{e_k^{n_k}} \;,
\end{eqnarray}
and this is similar as the case $f\ne 0$, the terms after decomposition have the same form as the original one, and one of the exponents $n_i$ decreases by $1$ in each term except $e_1$ whose exponent will increase by 1; the recursive relation is
\begin{equation}
G\PA{n_1,\cdots,n_N} = -\sum_{i=2}^N \frac{f_i}{f_1} G\PA{n_1+1,n_2,\cdots, n_i-1,\cdots,n_N}
\end{equation}
and we can repeat the decomposition until one of the $n_i$($i\ge2$) decreases to $0$.
\end{itemize}
So in each one of the two cases above, $G\PA{n_1,\cdots,n_N}$ can be reduced to the summation of $G\PA{n_1,n_2,\cdots, n_i =0,\cdots,n_N}$ which can not be decomposed any more, i.e. it is irreducible and the desired result.

If at least one exponent $n_{i_0}>0$ in Eq.~(\ref{Special}), without loss of generality, taking $i_0 = 1$ and $n_1>0$, then the element $e_1$ can be written as
\begin{equation}\label{Subeq}
e_1 = \frac{1}{f_1}  \PA{f - \sum_{i=2}^N f_i e_i} \;,
\end{equation}
we can substitute Eq.~(\ref{Subeq}) into Eq.~(\ref{Special}) to eliminate $e_1$:
\begin{equation}
\prod_{i=1}^{N} e_i^{n_i}  = \PB{\frac{1}{f_1}  \PA{f - \sum_{j=2}^N f_j e_j}}^{n_1} \prod_{i=2}^{N} e_i^{n_i} = \sum_{k} f'_k \prod_{i=2}^{N} e_i^{n_{ki}}\;,
\end{equation}
and now the final expression only involves $\{e_i\}_{(2\le i \le n)}$ which are linear independent, and it is irreducible.

To complete the proof, we will make the induction on $N$, i.e.~the number of elements in $\PC{e_i}$. It is trivial that this is valid for $N=1$, and now, assuming that it is also valid for $N = 1, 2, 3, \cdots, K$, we want to prove that it is also valid for $N = K+1$.

If $\PC{e_i}_{1\le i\le K+1}$ are linear independent, i.e.~irreducible, then there is no need for the decomposition; otherwise, there will be $M+1$($M\le K$) elements from $\PC{e_i}$ which are not linear independent, but any $M$ elements are linear independent. Without loss of generality, we take these elements as $\PC{e_i}_{1\le i\le M+1}$:
\begin{equation}
\prod_{i=1}^{K+1} e_i^{n_i} = \prod_{i=1}^{M+1} e_i^{n_i} \prod_{j=M+2}^{K+1} e_j^{n_j}\,.
\end{equation}
Then according to the special case we have considered in Eq.~(\ref{Special}), we have
\begin{equation}
\prod_{i=1}^{M+1} e_i^{n_i} = \sum_{j} f_j \prod_{i=1}^{N_j} e_{k_{ji}}^{n_{ji}} \;,
\end{equation}
where all $N_j \le M$, so
\begin{equation}
\prod_{i=1}^{K+1} e_i^{n_i} = \PA{\sum_{j} f_j \prod_{i=1}^{N_j} e_{k_{ji}}^{n_{ji}}} \prod_{m=M+2}^{K+1} e_m^{n_m}
= \sum_{j} f_j \PA{\prod_{i=1}^{N_j} e_{k_{ji}}^{n_{ji}} \prod_{m=M+2}^{K+1} e_m^{n_m}} \,.
\end{equation}
Since $N_j+(K-M) \le K$, i.e. the number of elements in $\PC{e_{k_{ji}}}_{1\le i\le N_j} \cup \PC{e_m}_{M+2\le m\le K+1}$ in the right hand side(rhs) is less than $N=K+1$, according to the assumptions, we have the following decomposition:
\begin{equation}
\prod_{i=1}^{N_j} e_{k_{ji}}^{n_{ji}} \prod_{m=M+2}^{K+1} e_m^{n_m} = \sum_{k} f'_k \prod_{i=1}^{N'_k} e_{k'_{ki}}^{n'_{ki}} \;,
\end{equation}
with each term in the rhs irreducible, so we get the decomposition for $N=K+1$:
\begin{equation}
\prod_{i=1}^{K+1} e_i^{n_i} = \sum_{k,j} f_j f'_k \prod_{i=1}^{N'_k} e_{k'_{ki}}^{n'_{ki}} \;.
\end{equation}
Since each term in r.h.s. is irreducible, the proof is done.

The procedure also gives us a method to perform the decomposition. We will give an implementation in \textsc{Mathematica}, i.e. the generalized \Apart function: \APart\!.

\section{An Implementation in \textsc{Mathematica}}
The basic functions in the package are:
\begin{itemize}

\item \texttt{\$Apart[expr,\{x,y,z,...\}]} \par
\texttt{expr} can be any form in Eq.~(\ref{General}), \texttt{\{x,y,z,...\}} are the corresponding n-variate polynomial variables, and \texttt{\$Apart[expr,\{x,y,z,...\}]} will perform the decomposition on \texttt{expr} to give the irreducible forms, which are expressed with the function \texttt{\$ApartIR}.

\def\subs#1#2{$\texttt{#1}_{\texttt{#2}}$}
\item \texttt{\$ApartIR[expr,\{x,y,z,...\},\{\subs{e}{1},\subs{e}{2},...,\subs{e}{N}\},\{\subs{n}{1},\subs{n}{2},...,\subs{n}{N}\}]} \par
where \texttt{expr} is actually the product of $\texttt{e}_i^{\texttt{n}_i}$, i.e.
\begin{equation}
\texttt{expr} = \prod_{i=1}^{\texttt{N}} \texttt{e}_i^{\texttt{n}_i} \,.
\end{equation}
We preserve the \{\subs{e}{1},\subs{e}{2},...,\subs{e}{N}\} and \{\subs{n}{1},\subs{n}{2},...,\subs{n}{N}\} for later use, because they will be used as the input parameters for \textsc{Fire}\cite{Smirnov:2008iw}. The irreducible form will be displayed as $\left\| \cdots \right\|$, and \texttt{\$RemoveApart} can be used to remove $\left\|\right.$ in $\left\| \cdots \right\|$.

\item \texttt{\$RemoveApart[expr]} \par
\texttt{\$RemoveApart} is used to remove the \textsc{Head} in \texttt{\$Apart} or \texttt{\$ApartIR}, and is defined as
\begin{center}
\texttt{\$RemoveApart[expr\_]:=expr/.\{\$Apart[x\_,\_]:>x,\$ApartIR[x\_,\_\_\_]:>x\}} \;.
\end{center}

\end{itemize}

We can take Eq.~(\ref{expect}) as the first concrete example:
\begin{equation}
\Apart\!\!\PB{\frac{1}{x (x + a) (x + y + b)},\texttt{\{x,y\}}} \Rightarrow \frac{\left\|\frac{1}{x (b+x+y)}\right\|-\left\|\frac{1}{(a+x) (b+x+y)}\right\|}{a} \;.
\end{equation}

As another relatively complicated case, we take
\begin{equation}\label{orginal}
\texttt{expr} = \frac{a}{(3a+b+c) (a+2b+d)^3 (a+4b+9e)} \,.
\end{equation}
If we only take $a$ and $b$ as the only polynomial variables, there are only two elements which can appear in the same irreducible expression, and we get the output as
\begin{eqnarray}\label{example}
\texttt{\$Apart[expr,\{a,b\}]} &\Rightarrow&
-\frac{(2 c-d) \left\|\frac{1}{(3 a+b+c) (a+2 b+d)^3}\right\|}{2 c-11 d+45 e}
+\frac{5 (4 c-9 e) \left\|\frac{1}{(3 a+b+c) (a+2 b+d)^2}\right\|}{(2 c-11 d+45 e)^2} \nn\\&&
-\frac{55 (4 c-9 e) \left\|\frac{1}{(3 a+b+c) (a+2 b+d)}\right\|}{(2 c-11 d+45 e)^3}
+\frac{121 (4 c-9 e) \left\|\frac{1}{(3 a+b+c) (a+4 b+9 e)}\right\|}{(2 c-11 d+45 e)^3} \nn\\&&
+\frac{2 (2 d-9 e) \left\|\frac{1}{(a+2 b+d)^3 (a+4 b+9 e)}\right\|}{-2 c+11 d-45 e}
+\frac{2 (4 c-9 e) \left\|\frac{1}{(a+2 b+d)^2 (a+4 b+9 e)}\right\|}{(2 c-11 d+45 e)^2} \nn\\&&
-\frac{22 (4 c-9 e) \left\|\frac{1}{(a+2 b+d) (a+4 b+9 e)}\right\|}{(2 c-11 d+45 e)^3} \;.
\end{eqnarray}
We can check the output with the original Eq.~(\ref{orginal}) using the code
\begin{eqnarray}
\texttt{dexpr} &=& \texttt{\$Apart[expr,\{a,b\}]} \nn\\
\texttt{comp} &=&\texttt{expr - (dexpr//\$RemoveApart)//Simplify}
\end{eqnarray}
The fact that \texttt{comp} gives zero indicates that the output \texttt{dexpr} is indeed identical with the original \texttt{expr}.

If take $c$ as a variable as well, we have
\begin{eqnarray}
\texttt{\$Apart[expr,\{a,b,c\}]} &\Rightarrow&
-\left\|\frac{1}{(3 a+b+c) (a+2 b+d)^3}\right\| \nn\\&&
+(9 e-2 d) \left\|\frac{1}{(3 a+b+c) (a+2 b+d)^3 (a+4 b+9 e)}\right\| \nn\\&&
+2 \left\|\frac{1}{(3 a+b+c) (a+2 b+d)^2 (a+4 b+9 e)}\right\|
\end{eqnarray}

More complicated examples can be found in \texttt{Example/Examples.nb} in the source code.

\begin{figure}[!h]
\centering\includegraphics[width=0.5\textwidth]{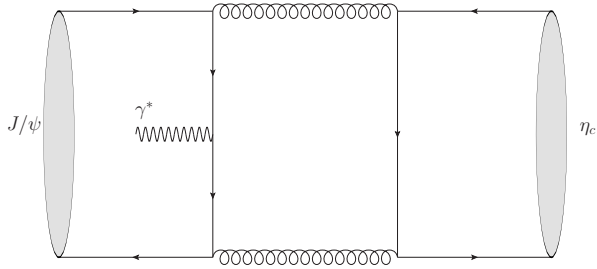}
\caption{A pentagon Feynman diagram for $e^+e^- \to\gamma^*\to J/\psi + \eta_c$\label{fig1}}
\end{figure}
Now, let us look at how to apply \APart to a specific feynman diagram from the process: $e^+e^- \to\gamma^*\to J/\psi + \eta_c$ which is shown in the Fig.~{\ref{fig1}}. After projecting the spin singlet and triplet for the charmonium $\eta_c$ and $J/\psi$ respectively with the spin projectors\cite{hep-ph/0211085} and performing the DiracTrace on the fermion chains, we get the amplitude for this diagram as
{\scriptsize
\begin{eqnarray}
\mathcal{A} = -\frac{16 i C_A C_F e g_s^4 \epsilon ^{\gamma \psi p_3p_4} \left((s-4) s k^2 m_c^2+k\cdot p_3^2+k\cdot p_4^2-(s-2) k\cdot p_3 k\cdot p_4\right)}{3 (D-2) m_c (s-4) s k^2 \left(k^2+k\cdot p_3\right) \left(4 m_c^2+k^2-2 k\cdot p_4\right) \left(2 s m_c^2+k^2-k\cdot p_3-2 k\cdot p_4\right) \left(k^2-k\cdot p_4\right)}
\end{eqnarray}
}where $p_3$ and $p_4$ are the momenta of $J/\psi$ and $\eta_c$ respectively, $\gamma$ and $\psi$ represent the polarizations of $\gamma^*$ and $J/\psi$ respectively, $k$ is the loop momentum, $m_c$ is the mass of the charm quark, and $s$ is defined by
\begin{equation}\label{sdef}
s \equiv \frac{Q^2}{4m_c^2} = \frac{(p_3+p_4)^2}{4m_c^2} \;.
\end{equation}

If we take $k^2$, $k\cdot p_3$ and $k\cdot p_4$ as the 3-variate polynomial variables, then the amplitude $\mathcal{A}$ has the same form as Eq.~(\ref{General}) after expanding the numerator, so we can perform the \APart operation on it:
\begin{equation}
\texttt{\$Apart[}\mathcal{A}, \texttt{\{}k^2, k\cdot p_3, k\cdot p_4\texttt{\}}\texttt{]}
\end{equation}

The output looks like: {\tiny
\makeatletter
\newcommand{\Vast}{\bBigg@{4}}
\makeatother
\allowdisplaybreaks[3]
\begin{eqnarray} &&
-\frac{4 i C_A C_F e g_s^4 \epsilon ^{\gamma \psi p_3p_4}}{3 (D-2) m_c^3 (s-4) (s-2) s}
\Vast[-4 \left\|\frac{1}{k^2 \left(k^2+k\cdot p_3\right) \left(4 m_c^2+k^2-2 k\cdot p_4\right)}\right\| m_c^2
+2 s \left\|\frac{1}{k^2 \left(k^2+k\cdot p_3\right) \left(2 s m_c^2+k^2-k\cdot p_3-2 k\cdot p_4\right)}\right\| m_c^2 \nn\\&&
-4 \left\|\frac{1}{k^2 \left(4 m_c^2+k^2-2 k\cdot p_4\right) \left(2 s m_c^2+k^2-k\cdot p_3-2 k\cdot p_4\right)}\right\| m_c^2
+2 s \left\|\frac{1}{\left(k^2+k\cdot p_3\right) \left(4 m_c^2+k^2-2 k\cdot p_4\right) \left(2 s m_c^2+k^2-k\cdot p_3-2 k\cdot p_4\right)}\right\| m_c^2 \nn\\&&
+2 (s-2) s \left\|\frac{1}{\left(k^2+k\cdot p_3\right) \left(4 m_c^2+k^2-2 k\cdot p_4\right) \left(k^2-k\cdot p_4\right)}\right\| m_c^2
+2 (s-2) s \left\|\frac{1}{k^2 \left(2 s m_c^2+k^2-k\cdot p_3-2 k\cdot p_4\right) \left(k^2-k\cdot p_4\right)}\right\| m_c^2 \nn\\&&
+2 (s-2) \left\|\frac{1}{k^2 \left(4 m_c^2+k^2-2 k\cdot p_4\right)}\right\|
+4 (s-2) \left\|\frac{1}{\left(k^2+k\cdot p_3\right) \left(2 s m_c^2+k^2-k\cdot p_3-2 k\cdot p_4\right)}\right\|
+(2-s) \left\|\frac{1}{k^2 \left(k^2-k\cdot p_4\right)}\right\| \nn\\&&
+(4-2 s) \left\|\frac{1}{\left(k^2+k\cdot p_3\right) \left(k^2-k\cdot p_4\right)}\right\|
+(2-s) \left\|\frac{1}{\left(4 m_c^2+k^2-2 k\cdot p_4\right) \left(k^2-k\cdot p_4\right)}\right\| \nn\\&&
+(4-2 s) \left\|\frac{1}{\left(2 s m_c^2+k^2-k\cdot p_3-2 k\cdot p_4\right) \left(k^2-k\cdot p_4\right)}\right\|\Vast]
\end{eqnarray}}
We can see that the 5-point integrals have been reduced to the 3-point or 2-point integrals, so the scalar integrals have been greatly simplified. Details of this example can be found in \texttt{Process/FC-43.nb} in the source code, where \texttt{43} is the sequence number of the corresponding diagram which has been generated by \textsc{FeynArts}\cite{Print-90-0144 (WURZBURG),hep-ph/0012260}.

\section{Application to Physical Loop Calculations}
The traditional method to compute cross sections for a physical process in perturbative quantum field theory involves generating
the amplitudes via Feynman diagrams and performing the dimensionally regularized loop integrals~\cite{'tHooft:1972fi}. Simplifications of the expressions are performed at the analytical level; there, an essential part is the reduction of these loop integrals to a small number of standard integrals. This step can be performed at the amplitude level for tensor integrals or, after contraction of Lorentz indices, at the level of interferences for scalar integrals. Considering the case of scalar integrals, integration by parts (IBP) identities~\cite{Tkachov:1981wb,Chetyrkin:1981qh} and Lorentz invariance (LI) identities~\cite{Gehrmann:1999as} may be used for a systematic reduction to a set of independent integrals, called master integrals (MI). The standard reduction algorithm by Laporta~\cite{laporta} defines an ordering for Feynman integrals, generates identities and solves the resulting system of linear equations. Alternative methods to exploit IBP and LI identities for reductions have been proposed~\cite{Smirnov:2005ky,Smirnov:2006tz,Gluza:2010ws,Schabinger:2011dz}; see also \cite{Lee:2008tj,Grozin:2011mt} and references therein. Public implementations of different reduction algorithms are available with the computer programs AIR~\cite{Anastasiou:2004vj}, FIRE~\cite{Smirnov:2008iw} and \textsc{Reduze}~\cite{Studerus:2009ye}.

As for the one-loop calculations, there are many automatic tools available to achieve the general one-loop amplitude, such as \textsc{FeynCalc}\cite{Mertig:1990an} and \textsc{FormCalc}\cite{hep-ph/9807565}, which are based on  the traditional Passarino-Veltman\cite{Passarino:1978jh,Denner:1991kt,Denner:2002ii,Denner:2005nn} reduction of Feynman graphs, which can be generated
automatically(FeynArts\cite{Print-90-0144 (WURZBURG),hep-ph/0012260} or QGRAF\cite{Nogueira:1991ex}). In order to produce numerical results, tensor
coefficient functions are calculated using \textsc{LoopTools}\cite{hep-ph/9807565}. See also Refs.~\cite{hep-ph/9602280,arXiv:0903.4665} and the references therein.

In the last few years, several groups have been working on the problem
of constructing efficient and automatized methods for the computation of
one-loop corrections for multi-particle processes. Many different interesting
techniques have been proposed: these contain numerical and semi-numerical
methods\cite{hep-ph/0402152,hep-ph/0508308,arXiv:0704.1835,arXiv:0708.2398}, as well as analytic approaches\cite{hep-ph/9409265,hep-ph/9403226,hep-th/0403047,hep-th/0406177} that make use of unitarity
cuts to build next-to-leading order amplitudes by gluing on-shell tree amplitudes\cite{hep-ph/0602178,hep-th/0611091}.
For a recent review of existing methods, see Refs.~\cite{arXiv:0707.3342,arXiv:0704.2798}.

In this section, we want to use the \APart and the \textsc{Fire}\cite{Smirnov:2008iw} package combined with \textsc{FeynArts}\cite{Print-90-0144 (WURZBURG),hep-ph/0012260} and \textsc{FeynCalc}\cite{hep-ph/9807565} to perform the one-loop calculations; here,
\textsc{FeynArts}\cite{Print-90-0144 (WURZBURG),hep-ph/0012260} and \textsc{FeynCalc}\cite{hep-ph/9807565} are used to generate the Feynman diagrams,
and to perform the DiracTrace respectively, the rest, such as tensor or scalar integral reductions, etc.~will be handled by the \APart and \textsc{Fire}\cite{Smirnov:2008iw} packages. We will concentrate on the next-to-leading-order corrections in $\alpha_s$ to double quarkonium production in $e^+e^-$ colliders, the basic procedure can be summarized as follows.
\begin{enumerate}
\item Use the \textsc{FeynArts}\cite{Print-90-0144 (WURZBURG),hep-ph/0012260} package to generate all Feynman diagrams for the partonic process: $e^+e^- \to \gamma^* \to c\bar{c} + c\bar{c}$.
\item Use \textsc{FeynCalc}\cite{hep-ph/9807565} to perform the \texttt{DiracTrace} and \texttt{SU(N)} color matrix trace.

\item Make an expansion in the relative momentum of the quark and the anti-quark in corresponding quarkonium and project out \texttt{S-}, \texttt{P-}, \texttt{D-},$\cdots$Waves.

\item Use the \texttt{\$Apart} to decompose the linear dependent propagators to independent ones.

\item Use the \textsc{Fire}\cite{Smirnov:2008iw} package to reduce the general loop integrals to master integrals (MI).

\item Process the final results, e.g.~to asymptotically expand the amplitudes or calculate the cross section.
\end{enumerate}

Taking the process $e^+e^- \to J/\psi + \eta_c$ as an example, for which there are large discrepancy between the NRQCD leading-order predictions and experimental data, and the important key step to resolve the discrepancies is that a large K factor of about $1.96$ has been found in the next-to-leading-order corrections in $\alpha_s$~\cite{Zhang:2005cha,Gong:2007db}.

The calculations at leading order can be found in the directory \texttt{Process/Tree}, and the next-to-leading-order calculations can be separated into several parts:
\begin{itemize}
\item \texttt{Process/Tree/FC-RN.nb} is used to calculate corrections from the counter-terms where the multiple renormalization is used.
\item \texttt{Process/FC.nb} is used to calculate the general loop corrections.
\item \texttt{Process/FC-Nf.nb} is used to calculate the corrections from the light quarks which are proportional to $\PA{\rm Nf-1}$.
\item \texttt{Process/Total.nb} will process the results generated from the above to give the numerical predictions or plots.
\end{itemize}
To compare with the results which are already present in other references, let us list some results which can be found in \texttt{Process/Total.nb}.

The asymptotically expanded amplitude at $s\gg 1$, with $s$ defined at Eq.~(\ref{sdef}), is
\begin{eqnarray}
\mathcal{A} &=& \mathcal{A}^{(0)} + \frac{\alpha_s}{\pi} \mathcal{A}^{(1)}+\mathcal{O}\PA{\alpha_s^2} \;,\nn\\
\frac{\mathcal{A}^{(1)}}{\mathcal{A}^{(0)}} &=& \frac{1}{72} \left[39 \ln^2s-9 (3+10\ln2) \ln s+300 \ln\!\frac{\mu}{m_c}+3 (195-53 \ln2) \ln2-2 \pi ^2-92\right] \nn\\
&&+ \frac{i \pi}{24} (-26 \ln s+30 \ln2+9) + \mathcal{O}\PA{\frac{1}{s}} \;.
\end{eqnarray}
This result agrees with Eq.~(6.4) of Ref.~\cite{Jia:2010fw} in which only the real part of asymptotic expansion is given, while our result also includes the imaginary part.

\begin{figure}[!h]
\centering
\includegraphics[width=0.45\textwidth]{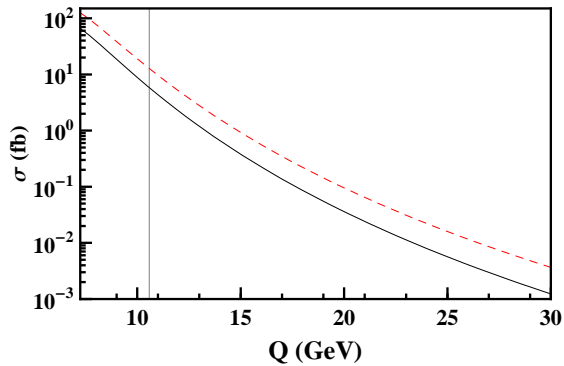}
\caption{Cross section for $e^+e^- \to J/\psi + \eta_c$ as function of the center-of-mass energy. The renormalization scale $\mu$ is set at half of the center-of-mass energy and $m_c=1.5$ GeV, the vertical line corresponds to $Q=10.6 {\rm GeV}$, solid line to $\sigma_{\rm LO}$ and dashed to $\sigma_{\rm NLO}$.\label{fig2}}
\end{figure}

We use the same input parameters as Ref.~\cite{Gong:2007db} to give the numerical results which are shown in Fig.~\ref{fig2} and Table~\ref{table}, these are consistent with Refs.~\cite{Zhang:2005cha,Gong:2007db}.
\begin{table}[!h]
\centering
\begin{tabular}{| ccccc | ccccc | ccccc | ccccc | ccccc | ccccc |}
\hline\hline
&& $m_c$(GeV) &&&&& $\mu$ &&&&& $\alpha_s(\mu)$ &&&&& $\sigma_{\rm LO}$(fb) &&&&& $\sigma_{\rm NLO}$(fb) &&&&& $\sigma_{\rm NLO}/\sigma_{\rm LO}$ &&\\
\hline
&& 1.5 &&&&& $m_c$ &&&&& 0.369 &&&&& 16.09 &&&&& 27.51 &&&&& 1.710 &&\\
\hline
&& 1.5 &&&&& 2$m_c$ &&&&& 0.259 &&&&& 7.94 &&&&& 15.68 &&&&& 1.975 &&\\
\hline
&& 1.5 &&&&& $Q/2$ &&&&& 0.211 &&&&& 5.27 &&&&& 11.14 &&&&& 2.113 &&\\
\hline
&& 1.4 &&&&& $m_c$ &&&&& 0.386 &&&&& 19.28 &&&&& 34.92 &&&&& 1.811 &&\\
\hline
&& 1.4 &&&&& 2$m_c$ &&&&& 0.267 &&&&& 9.19 &&&&& 18.84 &&&&& 2.050 &&\\
\hline
&& 1.4 &&&&& $Q/2$ &&&&&0.211 &&&&& 5.76 &&&&& 12.61 &&&&& 2.189 &&\\
\hline\hline
\end{tabular}
\caption{Cross sections with different charm quark mass $m_c$ and renormalization scale $\mu$, the input parameters are the same as Ref.~\cite{Gong:2007db} and $Q=10.6$GeV.\label{table}}
\end{table}

The method can be also used for calculations involving P-waves~\cite{Dong:2011fb}.

\section{Summary}
We have introduced a generalized \textsc{Mathematica} \Apart function, which will perform the decomposition on any linear dependent elements in $\mathcal{V}_{x}^*$ to reduce them to the irreducible form. The elements in $\mathcal{V}_{x}^*$ can be viewed as the corresponding propagators which involve loop momenta, and the decomposition will be useful when one tries to perform the loop calculations using the packages such as the \textsc{Fire} and the \textsc{Reduze}, which have implemented the integration by parts (IBP) identities and Lorentz invariance (LI) identities. A description of how to use this package, combined with \textsc{Fire}, \textsc{FeynArts} and \textsc{FeynCalc} packages, to do the one-loop calculations in double quarkonium production in $e^+e^-$ colliders is given, and the full source code for a specific process: $e^+e^-\to J/\psi + \eta_c$, from generating Feynman diagrams to the asymptotic expansion of amplitudes and numerical results, is also available.

\acknowledgments
The author wants to thanks Hai-Rong Dong, Wen-Long Sang and Prof. Yu Jia for many useful discussions. The research was partially supported by China Postdoctoral Science Foundation. Finally, The author would like to commemorate his beloved mother.

\end{document}